\documentclass{aa}
\usepackage[varg]{txfonts}
\usepackage{siunitx}
\usepackage{graphicx}
\usepackage{hyperref}

\bibpunct{(}{)}{;}{a}{}{,}

\hypersetup{ 
    colorlinks,
    linkcolor=blue,
    citecolor=blue
}

\newcommand{\pdv}[2]{\frac{\partial #1}{\partial #2}}
\newcommand{\funits}{\unit{kW.m^{-2}}}
\newcommand{\freq}{\unit{mHz}}
\newcommand{\Mm}{\unit{\mega\metre}}

\begin{document}

\title{Characteristics of acoustic-wave heating in simulations of the quiet Sun chromosphere}
\author{Elias R. Udn{\ae}s
        \inst{1,2},
        \and Tiago M. D. Pereira\inst{1,2}
        }
\institute{Rosseland Centre for Solar Physics, University of Oslo, P.O. Box 1029 Blindern, NO--0315 Oslo, Norway
\and
Institute of Theoretical Astrophysics, University of Oslo, P.O. Box 1029 Blindern, NO--0315 Oslo, Norway}

\abstract 
{Understanding energy transfer through the chromosphere is paramount to solving the coronal heating problem. We investigated the energy dissipation of acoustic waves in the chromosphere of the quiet Sun using 3D radiative magnetohydrodynamic (rMHD) simulations. We analysed the characteristics of acoustic-wave heating and its dependence on height and magnetic field configuration. We find the typical heights where acoustic waves steepen into shocks and the frequencies and wavenumbers that most efficiently dissipate wave energy through this steepening. 
We combined a comprehensive large-scale analysis, spanning the entirety of the simulations for several solar hours, with a detailed view of an individual shock. We find that the flux of propagating acoustic waves correlates closely with viscous dissipation in the chromosphere above the temperature minimum. Acoustic waves with frequencies close to the acoustic cut-off frequency can efficiently heat the quiet Sun chromosphere at the plasma-$\beta = 1$ interface and play an important role in the chromospheric energy balance.}

\date{}


\maketitle

\section{Introduction}

To understand the hot corona, we must look to the more complex dynamics of the chromosphere, a cooler part of the atmosphere situated closer to the surface. With a higher mass density and greater radiative losses than the corona, the chromosphere needs more energy to be heated; it requires a heating rate approximately an order of magnitude higher than the corona. \citet{De-Moortel:2015aa} suggested that coronal heating is only a side effect of chromospheric heating, so the coronal heating problem is really a chromospheric heating problem. Still, heating of the chromosphere remains a major unresolved issue \citep{Carlsson:2019aa}.

The current understanding of chromospheric heating includes two main processes: heating by wave dissipation and by magnetic reconnection. The magnetic field may also provide heating without reconnection, from current dissipation due to ambipolar diffusion, or field line braiding. It is unclear how these processes play out in the chromosphere since we are limited both by spatial and temporal resolution of solar observations, and spectral data can be complex to interpret. Heating of the solar atmosphere by waves generated in the convection zone was predicted already in the 1940s \citep{Biermann:1946aa, Schwarzschild:1948aa, Schatzman:1949aa}, but this theory was partly discredited by high-resolution observations with the transition region and coronal explorer \citep[TRACE][]{Handy:1999aa} in work done by \citet{Fossum:2005aa,Fossum:2006aa}. Shortly after these studies, however, \citet{Wedemeyer-Bohm:2007aa} used a 3D simulation to argue that while waves may not single-handedly balance radiative losses in the solar atmosphere, they play an important role for the energy balance of the solar atmosphere.

Today, the frontier in solar research lies in the interpretation of solar observations with the help of numerical simulations. 
This combination allows for a detailed description of the plasma, can disentangle the formation of radiation, and helps us study phenomena that are limited by atmospheric seeing conditions (especially for wave analyses) or that go beyond the diffraction limit of telescopes. 3D radiative magnetohydrodynamic (rMHD) simulations of the solar atmosphere -- such as those produced with {\it Bifrost} \citep{Gudiksen:2011vu} -- are self-contained; they produce atmospheric waves without periodic drivers at the lower boundary, as is common in 1D experiments of waves (e.g. the works done by \citealt{Stein:1972aa}, \citealt{Leibacher:1982aa}, and \citealt{Carlsson:1992vg} to name a few).

3D rMHD simulations can contain a convection zone, chromosphere, and corona, and with self-consistent wave generation are a good tool to study wave propagation through the solar atmosphere. In the convection zone of the simulation, \textit{p}-mode waves are created. In the photosphere, the \emph{p}-modes occur as the fast MHD wave, since the local sound speed tends to be much larger than the Alfvén speed. The fast mode is subject to mode conversion and transmission when it crosses the $\beta=1$ layer \citep{Bogdan:2003aa,Vigeesh:2009aa} and converts to MHD slow waves, Alfvén waves, or surface refraction, which leads to downward-propagating waves. In the chromosphere, \emph{p}-modes can steepen into shocks that dissipate energy to heat the plasma in a non-linear way.

\begin{figure*}
    \sidecaption
    \includegraphics[width=12cm]{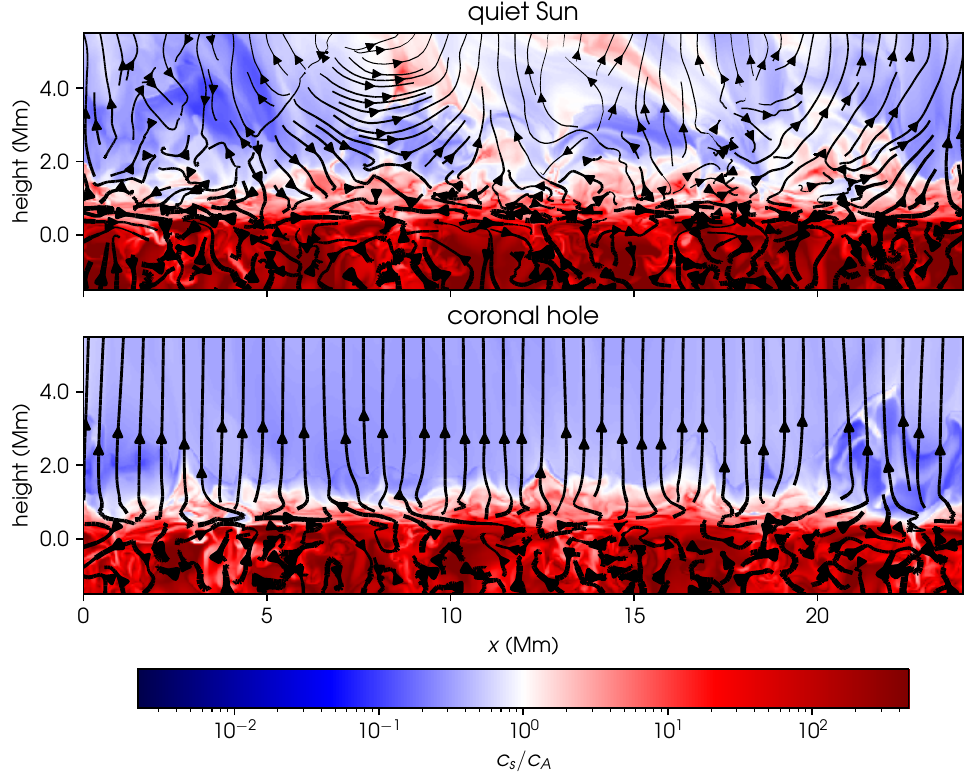}
    \caption{Ratio of sound speed to Alfvén speeds in $xz$ cuts taken from quiet Sun and coronal hole simulations. The simulation times are taken at 140 minutes and 40 seconds s and at $y$ coordinates \qty{3.10}{\Mm}. Overplotted are the magnetic field lines, with the thickness of the lines being scaled by the logarithm of the field strength, $\log_{10} \left\Vert \vec{B} \right\Vert$.}
    \label{fig:ca_cs}
\end{figure*}

According to the 1D hydrostatic model of \citet{Vernazza:1981aa}, the radiative losses in the chromosphere are 4.6~\funits. This value is often used as a canonical energy budget that has to be filled by either waves or magnetic energy. Different studies have produced different estimates of acoustic-wave fluxes, which do not always compensate for the canonical value of radiative losses. \cite{Fossum:2005aa} estimated acoustic fluxes around 0.4~\funits\ by combining TRACE observations with 1D dynamic simulations from RADYN \citep{Carlsson:1992vg}. \cite{Molnar:2021aa} also estimated acoustic fluxes of 1~\funits\ from RADYN simulations, but combined with Dunn Solar Telescope and Atacama Large Millimetre Array observations. On the other hand, \citet{Bello-Gonzalez:2009aa} and \citet{Bello-Gonzalez:2010aa} used spectropolarimetric observations from the Vacuum Tower Telescope \citep[see][and references therein]{Bello-Gonzalez:2008aa} to estimate an acoustic flux in the chromosphere of 4.6~\funits. More recently, \cite{Molnar:2023aa} used observations from the Interface Region Imaging Spectrograph \citep[IRIS,][]{De-Pontieu:2014aa} and inferred acoustic fluxes that are much lower than those in 3D rMHD \emph{Bifrost} simulations. A problem, as noted by \cite{Molnar:2023aa}, is that inferred acoustic fluxes are highly model-dependent. Moreover, the acoustic flux can be underestimated by the limited spatial resolution of observations, especially for studies of the chromosphere in radio wavelengths \citep{Molnar:2021aa}. 

Even if acoustic fluxes could be perfectly constrained, it remains unclear what portion of these fluxes in either observations or simulations contributes to atmospheric heating. Such analyses are obscured by acoustic-wave reflection, refraction, or transmission in the chromosphere. Complicating these analyses further is the mode conversion of magneto-acoustic waves at the $c_a = c_s$ layer. In this work, we studied how acoustic-wave flux dissipates to chromospheric heating in simulations of quiet Sun chromospheres. We find the characteristic frequencies and wavenumbers of acoustic waves being dissipated, as well as the typical altitudes of acoustic flux dissipation. Our analysis is limited to vertically propagating waves, an assumption often made in solar physics \citep[see e.g.][]{Bello-Gonzalez:2010ab}.

\section{Simulations}
In this work, we investigated acoustic-gravity waves in solar atmospheric simulations produced by the stellar atmosphere code {\it Bifrost} \citep{Gudiksen:2011vu}. The simulations provide access to the plasma parameters and we can readily analyse the state of the atmosphere. In our work, we studied two simulations with different magnetic field configurations: a quiet Sun simulation, {\tt cb24oi}; and a coronal hole simulation, {\tt ch024031\_by200bz005} (used in \citealt{Finley:2022aa}; that work contains a comprehensive description of the simulation). Both simulations cover the convection zone, photosphere, chromosphere, and corona, which has horizontal domains of \qtyproduct{24 x 24}{\Mm} and extends from \qty{2.5}{\Mm} into the convection zone to the corona \qty{14.3}{\Mm} above the surface. Both simulation boxes have $768^3$ grid points.

The quiet Sun simulation has a mean unsigned magnetic field at $z=0$ of \qty{4.1}{mT}. It was run for a total of 2 hours and 42 minutes after the simulation is relaxed. The coronal hole simulation has a mean unsigned magnetic field at $z=0$ of \qty{5.8}{mT}, and the simulation is run for 3 hours and 14 minutes after the simulation is relaxed. In the upper atmosphere, the magnetic field of the coronal hole is open and aligned in the $z$-direction. The magnetic field configurations of the simulations are shown in Fig.~\ref{fig:ca_cs}. The figure shows that not only are the field lines more vertical in the coronal hole, the magnetic field is also stronger in its upper atmosphere. 

These simulations are suited for wave studies due to their long duration of more than two hours, which gives a good frequency resolution for Fourier analyses. For both simulations we used 679 snapshots with a cadence of 10~s between each snapshot. The quiet Sun simulation was studied from 50 to 163 minutes, and the coronal hole simulation was studied from 113 to 226 minutes. The Nyquist frequency of the simulations is 50~\freq, and with a 113-minute time span we have a frequency resolution of $\approx \qty{0.15}{\freq}$. This frequency resolution is appropriate for, e.g. resolving individual peaks of the $p$-mode ridges in a power spectrum.

While MHD simulations are a great tool to study the solar atmosphere, there are limitations coming from assumptions that are made in the models. One artefact that is described in \citet{Carlsson:2016aa} is an unnaturally large amplitude of the global $p$ mode with a period of \qty{450}{s}. This is also discussed in \citet{Molnar:2023aa}, where signals were removed from low frequencies by Fourier filtering. Another challenge of {\it Bifrost} simulation is a too-dim UV spectrum \citep{Pereira:2013aa}. The too-weak Mg II UV lines are most likely due to too little mass loading of the chromosphere \citep{Hansteen:2023aa}, which can affect wave propagation in the upper chromosphere. For analyses of shockwaves, we also have limitations from the six-point stencil used by {\it Bifrost,} which, along with a quenching operator, suppresses small-amplitude perturbations and large gradients. Therefore, we cannot resolve a shockfront with fewer than six points in space, and shocks manifest as steep gradients instead of discontinuities.

\section{Analysis of MHD waves}

In MHD theory, three forces can act as restoring forces and transport waves through the medium: gas pressure, buoyancy from gravity, and magnetic tension. If one of them dominates, the MHD wave is a pure one-force mode (respectively, acoustic, gravity, or Alfvénic). Analysing the general three-force mode can be complex, and it is often a good approximation to neglect one component and focus on a two-force mode \citep{Stein:1974aa}.

For the quiet solar chromosphere, we can neglect magnetic tension and approximate MHD waves as an acoustic-gravity two-force mode. Its dispersion relation is thus

\begin{equation}
    k_z^2 = c_s^{-2}\left( \omega^2 - \omega_a^2 \right) - \left( \omega^2 - \omega_g^2 \right)\frac{k_h^2}{\omega^2},
    \label{eq:diagnostic}
\end{equation}
where $k_z$ and $k_h$ are the vertical and average horizontal wavenumbers, $c_s$ is the sound speed, and
$\omega_a$ and $\omega_g$ are the acoustic and internal gravity cut-off frequencies:

\begin{align}
    \omega_a &= \frac{\gamma g}{2 c_s} \, , \label{eq:omega_a} \\
    \omega_g &= \frac{\sqrt{\gamma - 1}g}{c_s} \, , \label{eq:omega_g}
\end{align}

where $\gamma$ is the adiabatic index.

\begin{figure}
    \resizebox{\hsize}{!}{\includegraphics{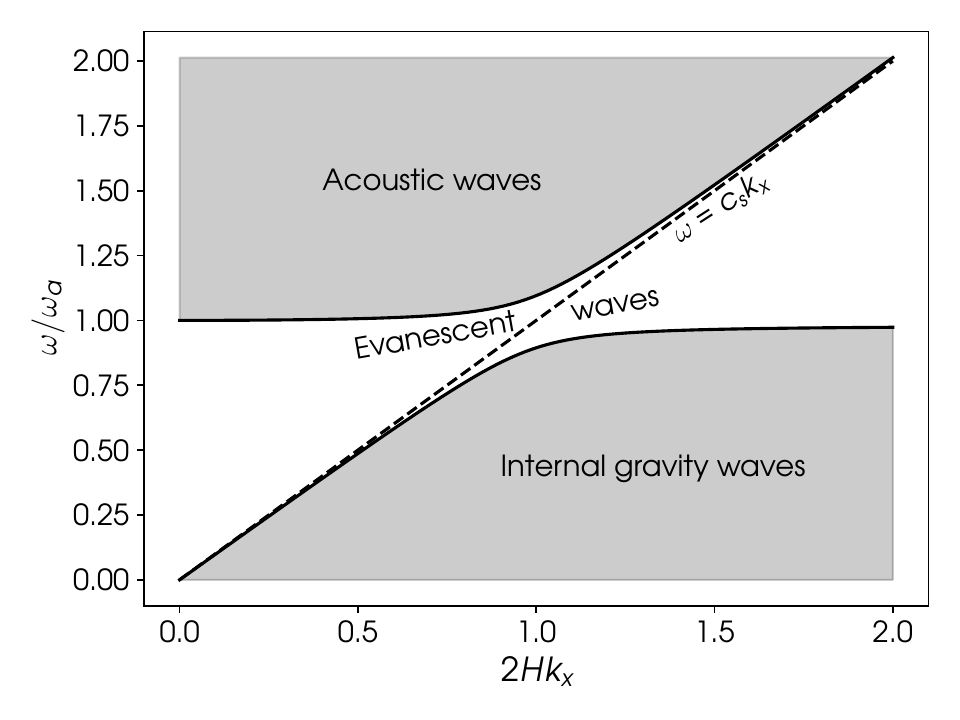}}
    \caption{Diagnostic diagram for acoustic-gravity waves in an isothermal atmosphere. Similar to \citet[][Fig. 53.1]{Mihalas1984found}, but with $\gamma = 5/3$, as used in our simulations.}
    \label{fig:diagnostic}
\end{figure}

\begin{figure*}
    \includegraphics[width=17cm]{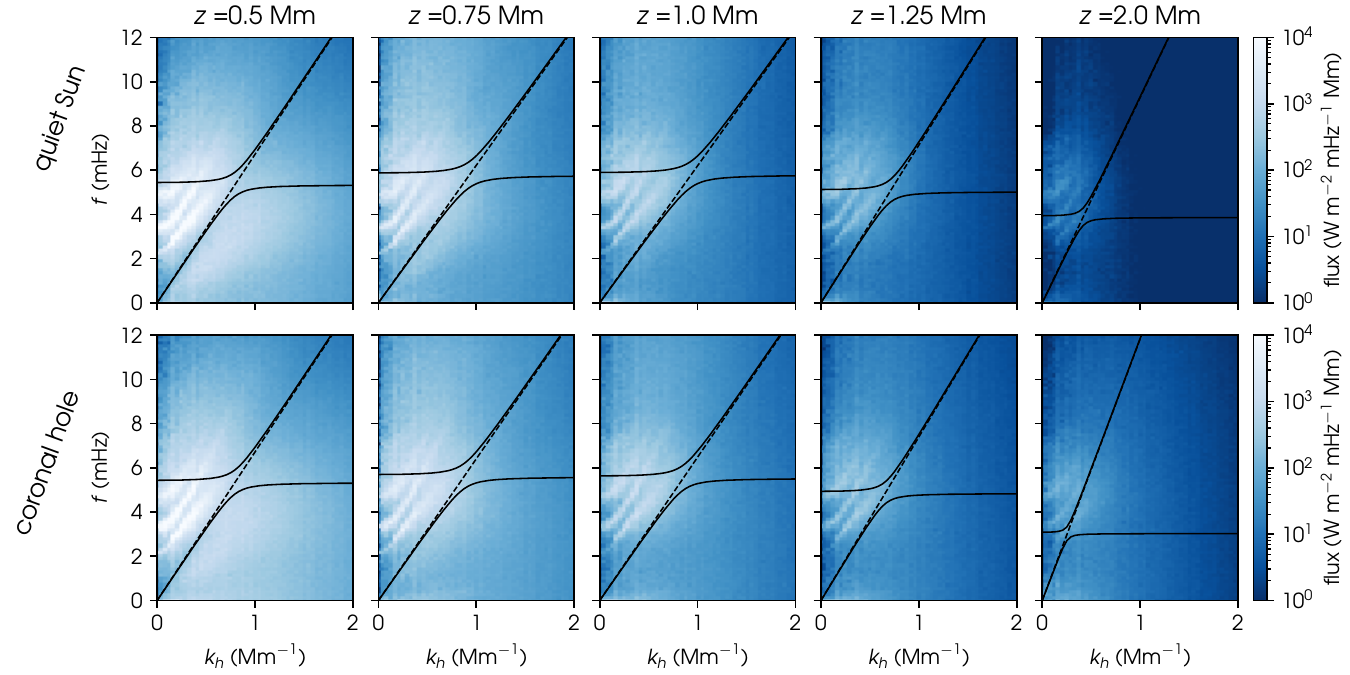}
    \caption{$k_h - \nu$ diagram of mechanical flux in quiet Sun and coronal hole simulations at five different heights in the atmosphere. The solid black lines are the dispersion relations calculated from time-averaged quantities at each height. The $x$-axes are divided by $2\pi$ and are therefore not in angular units.}
    \label{fig:power_spectra}
\end{figure*}

We show a dispersion relation diagram ($k_h$--$\nu$) in Fig.~\ref{fig:diagnostic} for the special case of an isothermal atmosphere with $\gamma=5/3$. The grey regions represent vertically propagating waves ($k_z^2 > 0$), while the white regions represent evanescent waves ($k_z^2 < 0$). In a non-isothermal atmosphere, the propagation boundaries are reduced \citep[see Fig.~3 of][]{Vigeesh:2017aa}. The diagram is also affected by radiative cooling (see Fig.~102.1 in \citealt{Mihalas:1978book} for the Newtonian cooling case).  

In this work, we estimated the energy deposition of waves in the upper atmosphere and considered upward-propagating waves. These could in principle be acoustic or internal gravity waves. From the dispersion diagram in Fig.~\ref{fig:diagnostic} one sees that only low-frequency gravity waves or higher frequency acoustic waves can propagate. Internal gravity waves are strongly damped by radiation in the lower atmosphere \citep{Mihalas:1982aa} and can be reflected back into the atmosphere beneath $\beta = 1$ \citep{Vigeesh:2017aa} or dissipated by non-linear breaking \citep{Mihalas:1981aa}. While internal gravity waves are important for the lower atmosphere, acoustic waves have a greater flux above \qty{\sim 700}{km} \citep{Straus:2008aa}. Therefore, we are more interested in the high-frequency acoustic waves' ability to heat the chromosphere.

Upward-propagating acoustic waves can dissipate energy in the upper atmosphere by shocks or by coupling to other MHD modes, where magnetic tension can no longer be ignored. The critical quantities governing which MHD mode occurs are the sound speed,

\begin{equation*}
    c_s = \sqrt{\gamma \frac{p}{\rho}} \,,
\end{equation*}
and Alfvén speed,
\begin{equation*}
    c_A = \frac{B}{\sqrt{\mu_0 \rho}} \,,
\end{equation*}
where $B$ is the magnetic field strength and $\mu_0$ is the vacuum magnetic permeability. The $c_s = c_A$ layer is the boundary between acoustic waves ($c_s > c_A$) and other MHD modes. The standard definition of plasma-$\beta$,
\begin{equation}
    \beta = \frac{n k_B T}{B^2/2\mu_0},
\end{equation}
can be written in terms of $c_s$ and $c_A$:
\begin{equation}
    \beta = \frac{2}{\gamma}\frac{c_s^2}{c_A^2}\,.
\end{equation}
We see that $\beta = 1$ is almost at the same height as $c_s = c_A$ when $\gamma = 5/3$. In the lower atmosphere where $c_s > c_A$, most propagating waves are acoustic. At low plasma-$\beta$ regions of the chromosphere, when $c_A > c_s$, the acoustic waves convert to either slow-mode MHD waves that travel with the sound speed, or continue as fast-mode waves traveling faster than the sound speed. \citet{Bogdan:2003aa} showed that it is only at the critical $c_s = c_A$ layer that most coupling between fast and slow MHD waves takes place. When $c_A > c_s$, most acoustic waves can develop into shocks, which are important for depositing energy in the higher layers. 

In Figure~\ref{fig:ca_cs}, we show an illustration of the distinct atmospheric regions by plotting the $c_s/c_A$ ratio from vertical cuts from the numerical simulations we employ. The region of equal sound speed and Alfvén speed is shown in white. The quiet Sun simulation is mainly dominated by pressure in the atmosphere beneath $z = \qty{1.5}{Mm}$. The quiet coronal hole simulation has a more magnetically dominated atmosphere, with the $c_s = c_A$ layer closer to \qty{1}{Mm} on average. 

A common tool to analyse waves is the Fourier transform. As part of a discrete Fourier transform (DFT) method, we use the amplitude of the wave with frequency $k$ as the magnitude of the DFT normalised by the number of data points in the input data. To obtain the Fourier power spectrum, we square the amplitude of the DFT and normalise it by the product of the frequency resolution and the wavenumber resolution. This way, we achieve a result that is independent of sampling frequency and spatial resolution.

The Fourier transform is used to calculate the coherence between two signals, which we define as
\begin{equation}
    K^2_{f, g} = \frac{C_{f,g}^2 + Q_{f, g}^2}{S_{f,f} S_{g,g}}\,,
    \label{eq:coherence}
\end{equation}
using the same terminology as \citet{Vigeesh:2017aa}. The terms are given as
\begin{align*}
    C_{f,g} &= \Re \left\{ \mathcal{F}(\vec{k}, \omega) \overline{\mathcal{G}(\vec{k}, \omega)} \right\} \,, \\
    Q_{f,g} &= \Im \left\{ \mathcal{F}(\vec{k}, \omega) \overline{\mathcal{G}(\vec{k}, \omega)} \right\} \,, \\
    S_{f,f} &= \left \langle \mathcal{F}(\vec{k}, \omega) \right \rangle^2\,.
\end{align*}

\section{Wave flux}

To study the propagation of wave energy, we used the mechanical flux term from \citet{Lighthill1987}:
\begin{equation}
    I = \rho_0 u^2 c_s \,,
\end{equation}
where $u$ is the speed of the fluid and $\rho$ is the gas density. By taking the Fourier transform of the velocity, $\hat{u} = \mathcal{F}(u)$, we obtained the mechanical flux as a function of frequency and wavenumber through the atmosphere:
\begin{equation}
    \mathcal{F}_m = \rho_0 \hat{u}^2 c_s\,
    \label{eq:mech_flux}
\end{equation} 
where $\hat{u}^2$ is the power  spectrum of $u$. The vertical velocity component $u = u_z$ was used to measure the vertical propagation of the mechanical flux, similarly to \citet{Molnar:2023aa}.

Prior to taking the Fourier transform of the vertical velocity signal, we normalised and apodised it. The normalisation was done by subtracting the mean, which removes the zero-frequency component of the Fourier transform. The apodisation was done with a Tukey window (also called a tapered cosine), which mitigates discontinuities at the edges of the signal. The coordinates in time and space were transformed to frequency $f$ with units millihertz, and wavenumber $k_x$ and $k_y$ with units per megametre. The 3D Fourier power of velocity was azimuthally averaged, transforming $k_x,k_y \rightarrow k_h$. This is a common technique used to more easily analyse Fourier powers in a $k_h - \nu$ diagram \citep{Jess:2023aa}. 

In this work, we focused on the interface region around the $\beta = 1$ layer. In geometric height, this layer is dynamic and corrugated (see Fig.~\ref{fig:ca_cs}, where $c_s = c_A$), and therefore we limited the analysis to the heights from the photosphere up to \qty{2}{\Mm}.

We calculated the mechanical flux for horizontal planes of the simulations using Eq.~\eqref{eq:mech_flux} and show it in Fig.~\ref{fig:power_spectra}. The vertical velocity power was multiplied by the average gas density and sound speed at each height. The lines in the figure are the dispersion relation of acoustic-gravity waves described by Eq.~\eqref{eq:diagnostic}. In the 1D linear theory outlined previously, the regions above and under these lines are where pressure waves and internal gravity waves propagate. The region between the lines is the evanescent region, where waves do not propagate. We calculated the dispersion relation using the average quantities at each layer. Due to the  temperature profile of the chromosphere, the average sound speed increases towards the temperature minimum, before declining again towards the hotter temperatures of the transition region. Therefore, the dispersion relation of acoustic waves has a maximal cut-off frequency at the temperature minimum.

In the $k_h - \nu$ diagram, we see strong wave power in the internal gravity wave region in the photosphere. This is expected, and it is the signature of convective motions. Higher up, the $p$-mode ridges are the only signature present. The velocity power at \qty{0.5}{\Mm} resembles the \ion{Fe}{i} power calculated from observations \citep[see, e.g. Fig.~5 of][which shows both $p$-mode ridges and a convective signature beneath the dispersion relation]{Kneer:2011aa}. The $p$ modes are the box modes of the simulation, and they lie mainly inside the acoustic and internal gravity cut-off frequency \citep[for simpler atmospheric models, this should be completely inside; shown in][]{Pinter:1999aa}. The overshooting ridges above the cut-off frequency appear because these 3D rMHD simulations are not an idealised isothermal plane-parallel atmosphere. They are leaking out of the convection zone and propagating through the chromosphere. These leaking \emph{p} modes above \qty{5.5}{mHz} are in fact believed to drive dynamic fibrils and some types of spicules \citep{Hansteen:2006aa, De-Pontieu:2007aa, Heggland:2007aa}.

\begin{figure}
    \resizebox{\hsize}{!}{\includegraphics{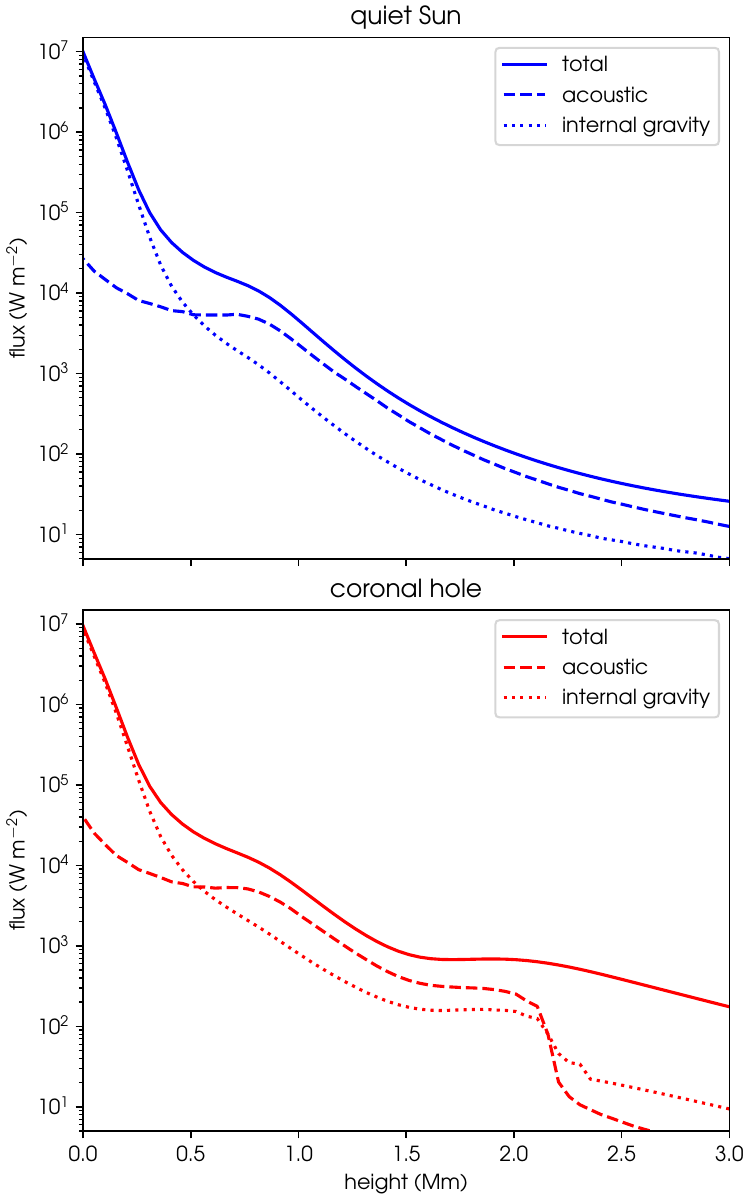}}
    \caption{Total integrated flux in quiet Sun and coronal hole simulations. Solid lines show the total mechanical flux; dashed lines show integrated acoustic flux; dotted lines represent integrated internal gravity flux. Contributions are determined from the dispersion relation that is calculated in Eq.~\eqref{eq:diagnostic} and shown in Fig.~\ref{fig:diagnostic}.}
    \label{fig:flux_total}
\end{figure}

To obtain the total vertical flux, we integrated the mechanical flux over frequency and wavenumber. We show the total, acoustic, and internal gravity fluxes for both simulations in Fig.~\ref{fig:flux_total}. To separate acoustic and internal gravity flux, we used the theoretical dispersion relations as a bounding filter. We calculated the acoustic flux by integrating above the dispersion relation, only including the propagating flux, and removing standing waves. This integration also removes the large global oscillation discussed in \citet{Carlsson:2016aa}.
The internal gravity flux was obtained by integrating in the region under the dispersion relation (labelled internal gravity waves in Fig.~\ref{fig:diagnostic}), and it should be taken as an upper bound since the group velocity of internal gravity waves is lower than the sound speed. Since our 3D simulations do not follow all assumptions used for the dispersion relation in Eq.~\eqref{eq:diagnostic}, these fluxes are not totally accurate and should be taken as proxies for the propagating and internal gravity-wave flux.

The flux terms in Fig.~\ref{fig:flux_total} are relatively high compared to studies with 1D simulations (e.g. \citet{Fossum:2005aa}). At $0.5$~\Mm, the acoustic flux is \qty{6}{\funits} for both simulations. This is high enough to balance the radiative losses of the chromosphere, at least the $4.6~\funits$ estimate of \citet{Vernazza:1981aa}. Still, this does not tell us that waves are solely responsible for heating a quiet Sun or coronal hole chromosphere. First, the acoustic flux presented here is a proxy. Second, different physical mechanisms can reduce the flux without leading to dissipation into heat as follows: wave reflection of acoustic waves back to the photosphere, mode conversion into different kinds of waves that travel into the corona, and wave damping by radiation. To heat the chromosphere, pressure waves have to develop into shocks that dissipate wave energy into heat. We analyse the energy deposition by looking at the heating terms in the simulations.

\begin{figure}
    \resizebox{\hsize}{!}{\includegraphics{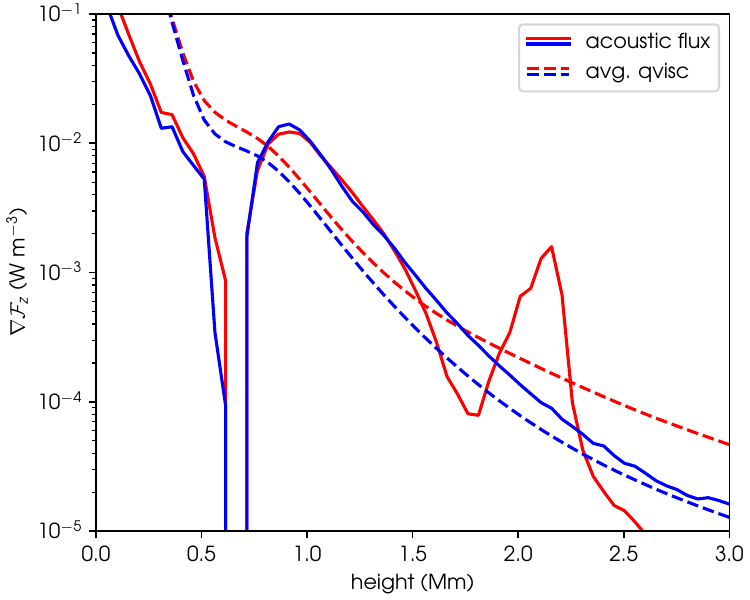}}
    \caption{Gradient of vertical acoustic flux (\emph{solid lines}) and time-averaged viscous dissipation (\emph{dashed lines}) in simulations of  the quiet Sun (\emph{blue}) and coronal hole (\emph{red}).}
    \label{fig:flux_grad}
\end{figure}

\begin{figure*}
    \includegraphics[width=17cm]{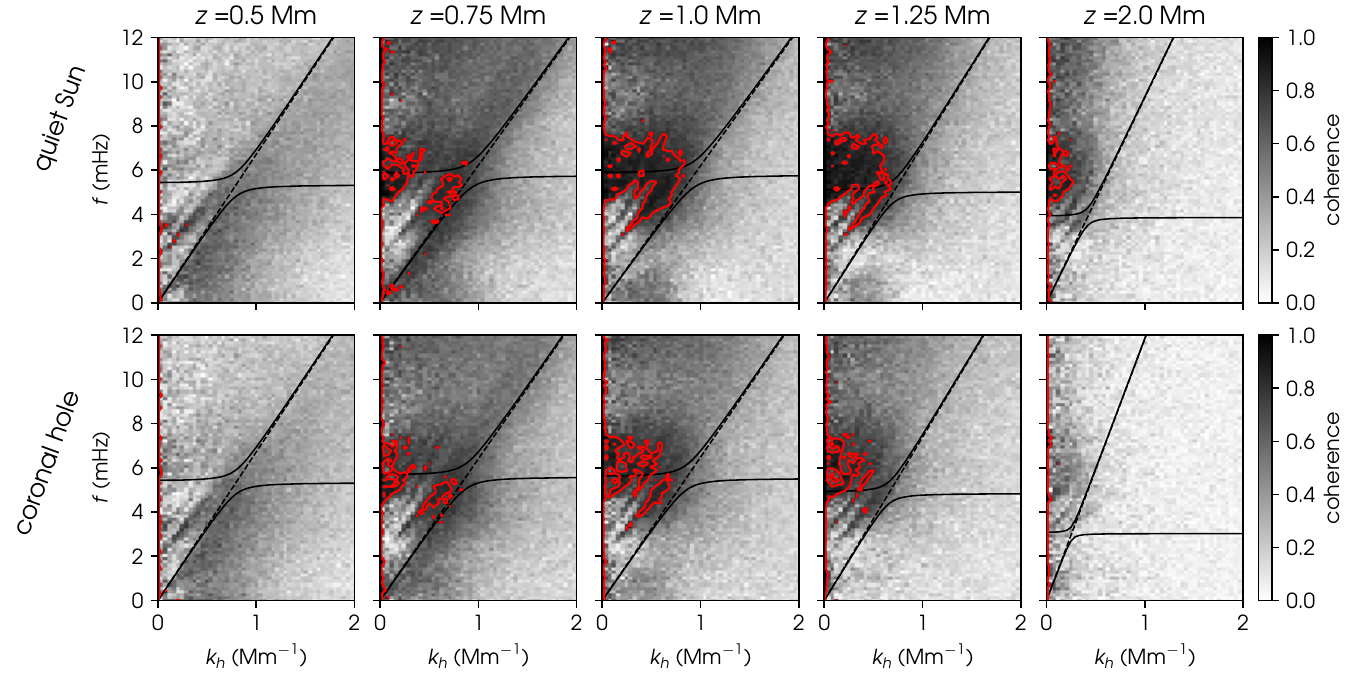}
    \caption{$k_h - \nu$ diagram of coherence $K_{m,q}$ between mechanical flux $\mathcal{F}_m$ and viscous dissipation $\mathcal{F}_q$. $K_{m,q}$, given by Eq.~\eqref{eq:coherence}, is calculated at five different heights in the simulated chromospheres. The solid black lines are the dispersion relations calculated from time-averaged quantities at each height. The red contour line outlines the $80\%$ confidence interval.}
    \label{fig:coherence}
\end{figure*}

\section{Energy dissipation from waves}

In our simulations, the wave flux decreases with height (Fig.~\ref{fig:flux_total}). This can be caused by many different processes. Waves can be reflected (e.g. by the transition region), or they may be refracted and change direction. However, the most relevant behaviour for this study is that waves can dissipate energy into heating of the chromosphere. To investigate the heat deposition by acoustic waves in the chromosphere, we analysed the heating terms in the simulations. 

Our {\it Bifrost} simulations have two main dissipative processes that heat the atmosphere: viscous and Ohmic (or Joule) dissipation \citep{Gudiksen:2005aa}. Viscous dissipation is the mechanism converting  wave flux to heat when acoustic waves develop into shocks, and it dominates Joule heating by about an order of magnitude in the mid chromospheres of the simulations. Our goal is to identify when viscous heating is brought on by waves and analyse what wave characteristics contribute to the heating.

If we take the height derivative of the acoustic flux, $F_{ac}$, we end up with the amount of acoustic-wave energy lost in the upward direction of the atmosphere. Energy is lost from  reflection or conversion from wave energy to heat, so the acoustic-wave heating should be smaller than this gradient:
\begin{equation*}
    \pdv{F_{ac}}{z} > \mathrm{acoustic~wave~heating} \,.
\end{equation*}
If high-frequency acoustic waves are the dominant source of viscous heating in the simulated chromospheres, then the equation above must be true.  

We calculated the derivative of the acoustic flux shown in Fig.~\ref{fig:flux_total}, and compared it with the average viscous heating per height in the simulations in Fig.~\ref{fig:flux_grad}. For intermediate heights (approx. 0.7--\qty{1.7}{Mm}), the acoustic-wave dissipation is greater than the viscous dissipation for both of the simulations. Around \qty{0.6}{Mm}, the gradient of acoustic flux is almost zero, which means the acoustic flux is not damped around the temperature minimum. The coronal hole simulation has a smaller loss of flux above \qty{1.5}{Mm}, before a spike appears at \qty{2}{Mm}, which is caused by a large gradient in sound speed in the transition region. This simulation has greater viscous dissipation higher up, which leads to larger flux. However, both the acoustic flux and its gradient decline rapidly above \qty{2}{Mm}, to the point where the acoustic flux is dropping below the viscous dissipation, meaning that other processes are driving the viscous dissipation. The quiet Sun simulation has a smaller acoustic flux gradient, but it remains above the viscous dissipation at higher layers. 

From Fig.~\ref{fig:flux_grad}, we can see that it is plausible that acoustic waves evolving into shockwaves are responsible for most of the viscous dissipation heating in the mid chromospheres of the quiet Sun and coronal hole simulations. A Fourier analysis of the viscous heating term, $\mathcal{F}_q,$ also shows high power in the $p$-mode ridges observed in $\mathcal{F}_m$ (see Fig.~\ref{fig:power_spectra}). While we do not show $\mathcal{F}_q$ here, we related the viscous heating to the mechanical flux by computing the coherence between $\mathcal{F}_m$ and $\mathcal{F}_q$ at each height using Eq.~\eqref{eq:coherence}. Calculating the $80\%$ coherence threshold, which is common in solar observations \citep{Jess:2023aa}, we obtain the characteristic frequencies and wavenumbers of wave heating, which we show in the $k_h - \nu$ diagram of Fig.~\ref{fig:coherence}. The figure shows the coherence above the $80\%$ confidence limit, which happens for intermediate heights only (\qtyrange{0.75}{1.25}{Mm}). 

acoustic-gravity waves that contribute to viscous heating have relatively high frequencies above \qty{4}{\freq}, but there are no signs of acoustic-wave heating above \qty{8}{mHz}. None of the coherence is inside the region of internal gravity waves, which is  expected. However, there are regions with high coherence below the acoustic cut-off frequency. This is likely because we are using a dispersion relation for an isothermal atmosphere, while our 3D model has local variations in temperature (e.g. cool pockets of gas in the chromosphere with significantly lower sound speeds).

\section{A qualitative view of heating in a wavefront}

So far, we have looked at average wave energy deposition across hundreds of simulation snapshots. We also wanted to know, however, what happens at the level of individual events. Here, we take a detailed look at a single event where a vertically propagating acoustic wave evolved into a strong shockfront. 

\begin{figure*}
    \sidecaption
    \includegraphics[width=12cm]{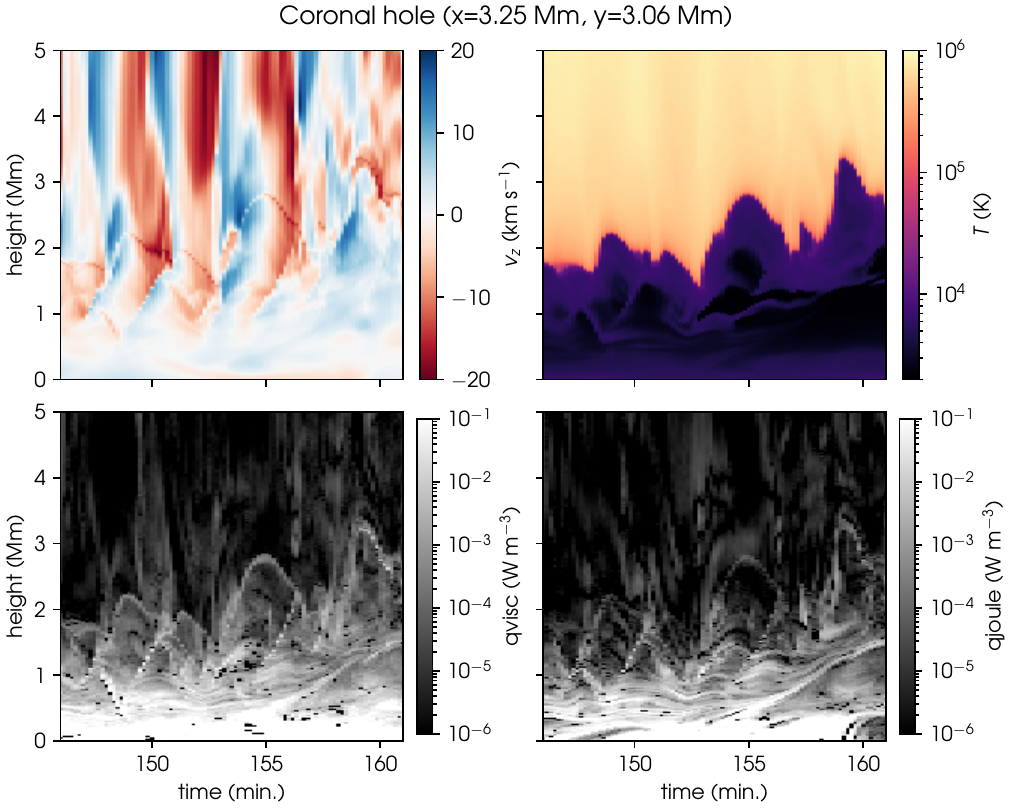}
    \caption{Vertical velocity, temperature, and dissipation coefficients in a column of the coronal hole simulation at $x=\qty{3.25}{\Mm}$ and $y=\qty{3.06}{\Mm}$.}
    \label{fig:overview_shock}
\end{figure*}

We selected the pixel at $(x=3.25~\mathrm{\Mm}, y=3.06~\mathrm{\Mm})$ from the coronal hole simulation and identified several vertically propagating wavefronts that developed into shocks in the chromosphere. In Fig.~\ref{fig:overview_shock}, we show the column's vertical velocity, temperature, and dissipation coefficients for 15~mins of simulated time. Wavefronts can be seen in all panels of the plot, but they are most identifiable in the vertical velocity. In the vertical velocity plot, we see gradients that propagate from the photosphere into the chromosphere and often reflect downwards off the transition region. These waves have typical periods of 3--4~min and come from the \emph{p} modes of the simulation.

We analysed a shock that developed at $t=149$~min in depth; this is visualised in Fig.~\ref{fig:trace_shock}. The shockfront is shown in vertical velocity in panel (a) of Fig.~\ref{fig:trace_shock}, with the location of the shockfront depicted by the solid black line. The dashed lines trace a particle propagating upwards with the local fast-mode speed or with the local sound speed. When the shock passes through the chromosphere, the gas density decreases, which leads to two things. First, the wave increases in amplitude due to energy conservation. The increased amplitude makes the wave develop into a strong shock and leads to wave heating, which can be seen in panels (c) and (e). Second, the magnetic pressure becomes comparable to the plasma pressure ($\beta \approx 1$). Therefore, the magnetic field becomes important in wave propagation. We see this in panel (a): the wavefront is supersonic; its speed is very close to the fast-mode speed from \qty{0.9}{\Mm} until \qty{1.4}{\Mm}. 

From the time when the shock starts to form at 149~min and 10~s, the dissipation coefficients increase as the shock travels upwards and becomes stronger. After 40~s, the integrated viscous dissipation reaches its peak (Fig.~\ref{fig:trace_shock} panel (c)) with a heating rate of \qty{4}{kW.m^{-2}}, before the shock decreases. When the shockfront passes $z=1.4$~\Mm, it reflects downwards towards the photosphere. 

In this example, as seen in panels (c) and (d) of Fig~\ref{fig:trace_shock}, both viscous and Ohmic dissipation are enhanced in the shock, but viscous dissipation is much higher. This is expected from friction arising in the shockfront. Integrating over the shockfront, we find that the maximum viscous dissipation is about $6.3$ times higher than the Ohmic dissipation. We find a similar scenario by looking at other individual events (not shown), where viscous dissipation often culminates above \qty{10}{kW.m^{-2}}. The heating in these events is of the same order of magnitude as radiative losses inferred from quiet Sun observations: \qty{4.5}{kW.m^{-2}} in \citet{Diaz-Baso:2021aa} or \qty{2.8}{kW.m^{-2}} in \citet{da-Silva-Santos:2024aa}. The detailed view from individual events is therefore consistent with the average flux results of the previous section, in particular those shown in Fig.~\ref{fig:flux_grad}, and it shows that shocks can deposit significant amounts of energy in the chromosphere. To find out exactly how much energy is deposited by all shocks, we would require a systematic accounting of shocks over all locations, which is beyond the scope of this work.

\section{Discussion}

We studied wave fluxes and energy deposition in 3D rMHD simulations, both from a global perspective and a detailed perspective of individual shocks. Estimating acoustic fluxes and wave dissipation from these complex numerical experiments is fraught with uncertainty, and so we adopted several assumptions to make the problem more tractable. 

We adopted the dispersion relation of an isothermal atmosphere, while the simulations are dynamic and inhomogeneous. We approximated the wave fluxes in Eq.~\eqref{eq:mech_flux} by using the horizontal averages of the gas density and sound speed. Using our dispersion relation implies adiabatic wave propagation, an assumption that can be troublesome in the presence of vertical temperature gradients, which increase the cut-off frequency \citep{Vigeesh:2017aa}. However, the effects of radiation counteract vertical temperature gradients, pushing the propagation closer to the adiabatic case \citep{Mihalas:1978book,Khomenko:2008aa}. For example, using the Newtonian radiation field described in \cite{Souffrin:1966aa} and \cite{Mihalas:1978book}, with solar conditions at the temperature minimum ($T \approx 2700$~K and $\gamma = 5/3$), the acoustic cut-off frequency for an instantly relaxed radiation field is \qty{3.6}{mHz}, while an adiabatic atmosphere has an acoustic cut-off frequency of \qty{4.6}{mHz} (a Newtonian radiation field mimics optically thin radiation). With optically thick losses, radiation is not effectively altering the dispersion relation of \citet{Heggland:2011aa}. However, inclined magnetic fields allow the propagation of low-frequency acoustic waves in the chromosphere \citep{De-Pontieu:2004aa,Hansteen:2006aa,Jefferies:2006aa}.
Therefore, an accurate dispersion relation for wave propagation is not feasible, and the adiabatic propagation is a fair approximation for our purposes. Propagation at lower frequencies will happen above inclined magnetic field concentrations, and we reiterate that our flux estimates are a lower bound.

Wave flux at high frequencies can differ significantly between 3D rMHD simulations run with different codes by up to two orders of magnitude \citep{Fleck:2021aa}. Therefore, high-frequency flux should be approached with care. However, in our case its contribution to the total flux is small because most of the chromospheric flux is concentrated around \qty{5}{mHz}.

We find that internal gravity waves have very low power in the upper photosphere and chromosphere and are effectively damped in these higher layers, which is consistent with results of earlier works \citep[e.g.][]{Mihalas:1981aa, Mihalas:1982aa, Vigeesh:2017aa}. Non-linear wave breaking of internal gravity waves \citep{Mihalas:1981aa,Vigeesh:2017aa} is dominant in the lower and middle chromospheres, turning propagating internal waves into turbulence. At the height of wave breaking, the maximal vertical velocity of internal waves is a few kilometres per second; that is,  much lower than the group velocity of acoustic waves \citep{Mihalas:1981aa}. Damping at lower heights is shown in \citet{Mihalas:1982aa}, which identified radiative transfer effects that smooth the temperature variations from internal gravity waves, damping them from the low photosphere to the temperature minimum. The suppression of internal gravity waves shifts the peak of the acoustic-gravity flux to higher frequencies, making high-frequency waves more relevant for chromospheric heating.

In Fig. \ref{fig:flux_grad}, we show that internal gravity waves are not as important as acoustic waves for chromospheric heating. The acoustic  and gravity-wave-heating regions correspond to the high-frequency part of the $p$-mode ridges seen in Fig.~\ref{fig:power_spectra}. These leaked $p$ modes come from the convection zones of the simulations, travel vertically, and develop into shocks that dissipate wave energy in the chromosphere. Since most of the coherence is located at heights of around \qty{1}{Mm} (Fig.~\ref{fig:coherence}), acoustic-wave heating takes place in the middle of the chromosphere. Above this region, there may still be wave heating, but it will be from waves that have the magnetic field as a restoring force; that is, from mode conversion.

We see maximal coherence between acoustic flux and viscous heating around \qty{1}{Mm}, where the simulations typically have the $c_s = c_A$ divide that is seen in Figure~\ref{fig:ca_cs}. This layer is also where fast and slow magneto-acoustic-gravity waves decouple \citep{Bogdan:2003aa}. Under this layer, where pressure dominates, the fast and slow magneto-acoustic-gravity waves are essentially the fast-mode acoustic wave. The reason for the high coherence between acoustic flux and heating at this layer is two-fold. First, the acoustic waves develop into shocks at this height, which is essential to producing viscous heating. Second, above this height, we have mode conversion of the acoustic waves. The coronal hole atmosphere has a stronger and more uniform magnetic field, which introduces magnetic field dependency to the acoustic waves lower down in the atmosphere. Therefore, the acoustic heating coherence shown in Fig.~\ref{fig:coherence} is smaller for the coronal hole simulation. Other reasons for lower acoustic-wave heating above the $c_s = c_A$ layer include the dissipation and reflection of acoustic shock waves. 

\begin{figure*}
    \includegraphics[width=17cm]{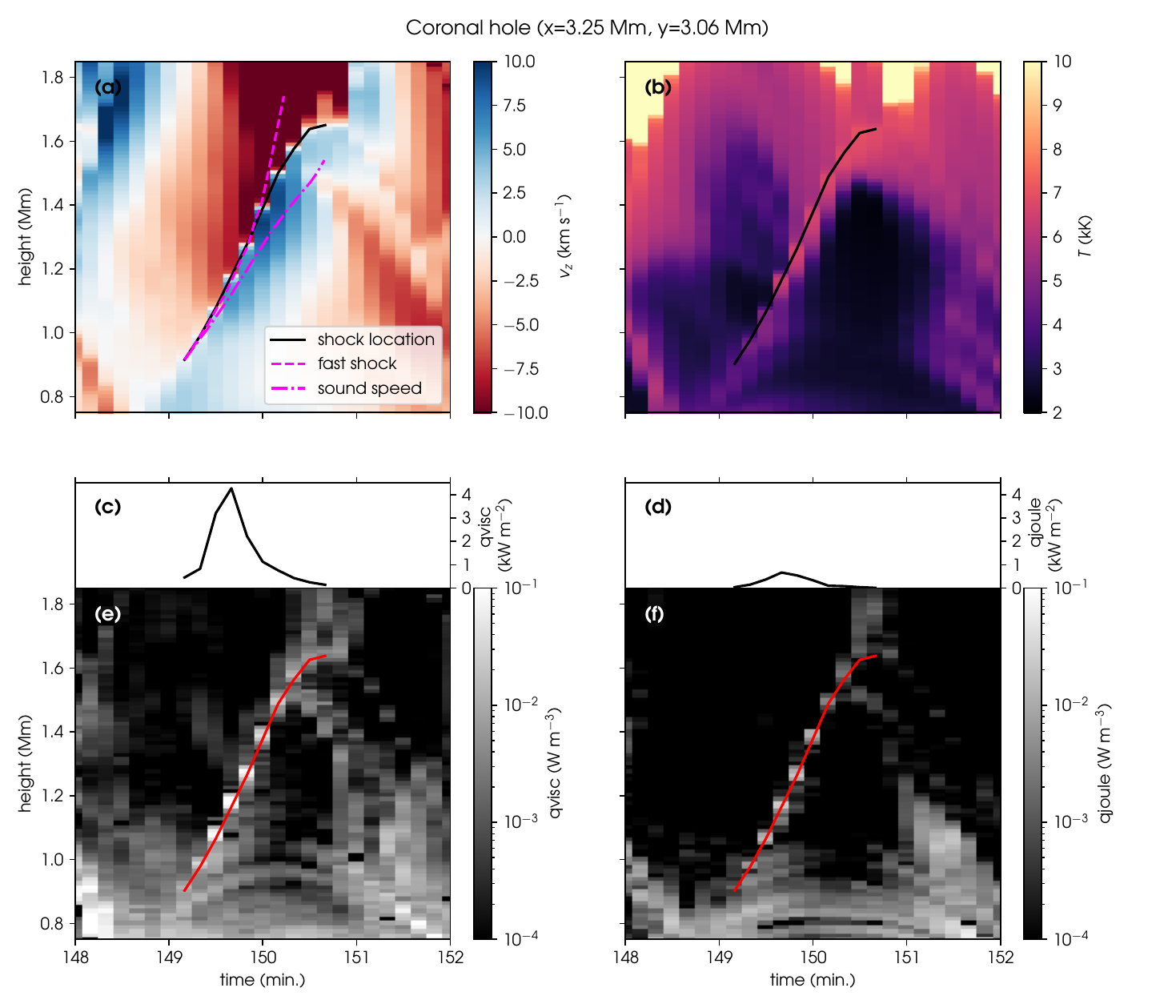}
    \caption{A column of the coronal hole simulation where a wave front develops into a shock. $x$-axes show simulation time in minutes, and $y$-axes show height above the surface in Mm. Panel (a) shows vertical velocity, (b) shows temperature, and (e) and (f) show the viscous and Ohmic (or Joule) volumetric dissipation coefficients. The lines of panels (c) and (d) are the volumetric dissipation coefficients integrated over the 5 grid points constituting the computational stencil of the shock front. The solid lines in panels (a), (b), (c), and (d) show the location of the shock front. The dashed and dash-dotted magenta lines in panel (a) traces particles travelling with the local fast-mode and sound speed $c_s$ respectively.}
    \label{fig:trace_shock}
\end{figure*}

In Fig.~\ref{fig:flux_total}, we see a rapid decline in acoustic flux around \qty{1}{Mm}, which is also seen as a peak in the gradient of acoustic flux at \qty{1}{Mm} in Fig.~\ref{fig:flux_grad}. Some of the decline in acoustic flux is due to mode conversion, but from Fig.~\ref{fig:flux_grad} we also see viscous dissipation that is comparable to the loss of acoustic flux. Looking at an individual shockfront, we find mode conversion from an acoustic shock to a fast magneto-acoustic shock above \qty{1}{Mm}, while the shockfront produced high heating rates in the chromosphere before reflecting off  the transition region (Fig.~\ref{fig:trace_shock}). Reflection of the fast shock can be due to the steep gradient in the Alfvén speed \citep{Khomenko:2012aa}.

It is difficult to study the viscous dissipation power $\mathcal{F}_q$ quantitatively. Because viscous dissipation is not a smooth function, its Fourier transform shows Gibbs phenomenon \citep{Wilbraham:1848} artefacts. However, we saw large regions with more than $\qty{80}{\%}$ coherence between acoustic flux and viscous dissipation, which we show in Fig.~\ref{fig:coherence}. We then used the coherence between acoustic-gravity flux and viscous dissipation as a proxy to identify the locations and frequencies where acoustic-wave heating takes place. We find that acoustic-wave heating is most dominant in the region where the plasma transitions from being pressure-dominated to magnetically dominated. The simulations also show acoustic-wave heating centred around the acoustic cut-off frequency at spatial scales bigger than \qty{1}{Mm}. The coronal hole simulation displays a lower acoustic-wave-heating coherence than the quiet Sun simulation, which indicates that wave propagation already depends on the magnetic field at heights of \qty{1}{Mm}. Magnetic field effects on a shockfront can also be seen in Fig.~\ref{fig:trace_shock}. Here, the shock propagated faster than the sound speed above a height of \qty{1}{Mm}, where it propagated as a fast magneto-acoustic shock. Above \qty{1.4}{Mm}, the shock was reflected from the transition region towards the lower atmosphere. 

Our analysis rests on two main limitations: how accurately we can estimate shock heating from simulations, and the fact that we analyse only vertically propagating waves. 

Sharp shockfronts are predicted to have a width of only a few mean-free paths \citep{Priest:2014aa}. A 3D simulation is not realistically able to resolve the shockfront, because grid sizes are not as small and the numerical stencils need to span several grid points. In a 3D simulation, we see shocks as steep gradients with a width comparable to the numerical stencil instead of being a jump over an infinitesimal thickness. In {\it Bifrost}, the numerical diffusion \citep[diffusive and quench operator; see][]{Gudiksen:2011vu} spreads the immediate heating from the shockfront out over five-to-six grid points. When we recovered the heating from the shockfront, we therefore counted contributions from the grid points that constituted the numerical stencil around the shockfront. If anything, this leads to a smoothing of the estimated heat deposition, and our values should be a lower bound, which has a minimal effect on the results.

While we chose to only study vertically propagating acoustic waves to make the analysis feasible, it is also a reasonable assumption to make when studying energy transport through the chromosphere. Acoustic waves have smaller horizontal amplitudes than vertical ones in the lower atmosphere \citep{Bello-Gonzalez:2010ab}, and horizontal amplitudes are more important for internal gravity waves \citep{Mihalas:1982aa}, which do not transport much flux into the chromosphere. \citet{Fossum:2006aa} discussed that a spherical wavefront that is excited in the deeper atmosphere will be refracted to a more planar wave due to the change in the sound speed, and it will travel in the direction of the gradient of the sound speed, which for most cases is in the vertical direction. Therefore, we believe that only accounting for vertically propagating waves means that our values should be a lower bound, and so this does not affect our results.

Studies using 1D simulations \citep{Fossum:2005aa,Fossum:2006aa} and 1D hydrostatic models \citep{Sobotka:2016aa} combined with observations have found that acoustic-wave heating is not high enough to balance radiative losses. Still, these studies discuss that limitations in the spatial resolution can negatively affect flux calculations, a point that is reinforced by findings in \citet{Yadav:2021aa}, which found that current observations give flux estimates roughly a factor of three too small. \citet{Sobotka:2016aa} concluded that acoustic waves may still may become the main contributor to chromospheric heating in certain chromospheric regions. \citet{da-Silva-Santos:2024aa} also found that non-LTE inversions can underestimate acoustic fluxes in the chromosphere and overestimate radiative losses. Here, we did not calculate the proportion of acoustic flux that is dissipated to heat in the chromosphere, and therefore we make no claim about the proportion of heating that comes from acoustic waves in the quiet chromosphere. However, we see that individual events provide substantial heating where acoustic-wave heating dominates. This suggests that acoustic-wave heating must be an important component in the energy balance of the quiet chromosphere.

\section{Conclusions}
In this work, we looked into the role of acoustic-wave heating in simulations of quiet chromospheres. Analysing two rMHD simulations with quiet magnetic fields, we obtained estimates of acoustic flux and wave-heating characteristics in the photosphere and chromosphere of the simulations. One simulation had a magnetic field configuration typical of the quiet Sun, and the other had that of a coronal hole. The open magnetic field of the coronal hole gave slightly different wave heating characteristics than for the quiet Sun.

Using a simple acoustic-gravity dispersion relation, we separated the acoustic flux into two components: acoustic and internal gravity waves. This approach worked well for the lower atmosphere of the simulations, where the plasma is pressure-dominated, and the sound speed is higher than the Alfvén speed. In the chromospheres we find that most wave flux is acoustic; the internal-gravity flux is quickly damped in the photosphere and is weaker in the chromosphere (above \qty{0.5}{Mm}). $k_h - \nu$ diagrams show strong chromospheric flux around the $p$-mode ridges of the simulations, with frequencies from \qtyrange{3}{6}{mHz} and horizontal wavenumbers from \qtyrange{0}{1}{Mm^{-1}}. While most of these power ridges are underneath the acoustic cut-off frequency, in the evanescent region, we find that a large fraction of the flux is above the acoustic cut-off frequency in the simulated chromospheres.

The heating from acoustic-gravity waves is centred mainly at heights of around \qty{1}{Mm} for both simulations, which is the region where, on average, plasma-$\beta$ reaches unity. Acoustic heating is more pronounced in the quiet Sun simulation. In the coronal hole simulation, acoustic heating is less distinct due to the influence of the magnetic field on the chromospheric waves. A qualitative analysis showed a vertically propagating shockfront converted into a fast magneto-acoustic shock in the coronal hole chromosphere, before being reflected off the transition region. Together, these analyses highlight the importance of the magnetic field configuration on the propagation of acoustic-gravity waves across the chromosphere: a weak magnetic field facilitates the propagation of acoustic waves, and these waves dissipate most of their energy in the region where plasma pressure balances the magnetic pressure.

Finally, we compared the acoustic flux to the average viscous dissipation at different heights. At heights above \qty{0.75}{Mm}, we find that the energy lost by acoustic waves is more than enough to account for the heat generated by viscous dissipation of waves. This view is complemented by the analysis of single shock events, which shows that acoustic waves are the largest contributors to viscous dissipation in the simulated chromospheres. We conclude that acoustic-wave heating is an important mechanism in the chromospheric energy balance for the quiet Sun and coronal holes. Further work is needed to identify and measure the prevalence of shocks, so one can estimate the total energy deposited by acoustic waves and relate it to radiative losses and ultimately the heating of the chromosphere.

\begin{acknowledgements}
This work has been supported by the Research Council of Norway through its Centers of Excellence scheme, project number 262622. We kindly acknowledge the computational resources provided by UNINETT Sigma2 - the National Infrastructure for High Performance Computing and Data Storage in Norway.
\end{acknowledgements}

\bibliographystyle{aa}
\bibliography{references}

\end{document}